\documentclass[tradiabstract]{aa}

\usepackage{txfonts,epsfig,graphicx,natbib,url,twoopt}
\usepackage[breaklinks=true]{hyperref} 

\bibpunct{(}{)}{;}{a}{}{,}    
\newcommandtwoopt{\citeads}[3][][]{\href{http://adsabs.harvard.edu/abs/#3}%
                                        {\citealp[#1][#2]{#3}}}
\newcommandtwoopt{\citepads}[3][][]{\href{http://adsabs.harvard.edu/abs/#3}%
                                         {\citep[#1][#2]{#3}}}
\newcommandtwoopt{\citetads}[3][][]{\href{http://adsabs.harvard.edu/abs/#3}%
                                         {\citet[#1][#2]{#3}}}
\newcommandtwoopt{\citeyearrads}%
  [3][][]{\href{http://adsabs.harvard.edu/abs/#3}%
                                         {\citeyear[#1][#2]{#3}}}

\begin{document}

\twocolumn[{%
\vspace*{4ex}
\begin{center}
  {\Large \bf Temperature anisotropy and differential streaming of solar wind ions -- Correlations with transverse fluctuations}\\[4ex]          
  {\large \bf Sofiane Bourouaine$^{1}$,
              Eckart Marsch$^1$
              and
              Fritz M. Neubauer$^2$}\\[4ex]
  \begin{minipage}[t]{15cm}
        $^1$ Max-Planck-Institut f\"ur Sonnensystemforschung,
37191 Katlenburg-Lindau, Germany\\
        $^2$ Institut f\"{u}r Geophysik und Meteorologie, Universit\"{a}t
zu K\"{o}ln, Albertus-Magnus-Platz, K\"{o}ln, 50923, Germany\\

{\bf Abstract.} We study correlations of the temperature ratio (which is an indicator for
perpendicular ion heating) and the differential flow of the alpha particles with
the power of transverse fluctuations that have wave numbers between $0.01$
and $0.1$ (normalized to $k_p=1/l_p$, where $l_p$ is the proton inertial
length). We found that both the normalized differential ion speed,
$V_{\alpha p}/V_\mathrm{A}$ (where $V_\mathrm{A}$ is the Alfv\'en speed) and
the proton temperature anisotropy, $T_{\perp p}/T_{\parallel p}$, increase
when the relative wave power is growing. Furthermore, if the normalized
differential ion speed stays below 0.5, the alpha-particle temperature
anisotropy, $T_{\perp \alpha}/T_{\parallel \alpha}$, correlates positively
with the relative power of the transverse fluctuations. However, if
$V_{\alpha p}/V_\mathrm{A}$ is higher than 0.6, then the alpha-particle
temperature anisotropy tends to become lower and attain even values below
unity despite the presence of transverse fluctuations of relatively high
amplitudes. Our findings appear to be consistent with the expectations from
kinetic theory for the resonant interaction of the ions with
Alfv\'en/ion-cyclotron waves and the resulting wave dissipation.

\vspace*{2ex}
\end{minipage}
\end{center}
}]





\section{Introduction}

The in-situ measurements of protons and heavier ions in fast solar wind
revealed distinct non-thermal kinetic features, such as the core temperature
anisotropy and beam of the protons, or the preferential heating and
acceleration of alpha particles and other minor species (see, e.g., the
review of \cite{marsch2006} on this subject). Similarly, solar remote-sensing
observations of coronal holes, which are known as the sources of the fast solar
wind provided evidence via significant broadenings of ultraviolet emission
lines, for strong perpendicular heating of oxygen and heavy ions in the solar
corona (see, e.g., the review of \cite{cranmer2009}).

These observed kinetic features have usually been interpreted as signatures
of heating by ion-cyclotron waves
\citep{isenberg2001,marsch2001,hollweg2002,matteini2007}. Their origin
remains unclear, though, and is still a matter of debate. Some authors
suggested that a nonlinear parallel cascade (via parametric decay) of
low-frequency Alfv\'{e}n waves may ultimately generate ion-cyclotron waves,
since their numerical simulations showed that protons and heavy ions can be
heated perpendicularly by wave absorption through cyclotron resonance
\citep{araneda2008, araneda2009}. Moreover, it has been argued theoretically
that in low-beta plasma condition, the ion temperature anisotropy and the
preferential (stochastic) heating of heavy species could even be caused by
Alfv\'enic fluctuations with frequencies well below the local ion-cyclotron
frequency \citep{wu2007, bourouaine2008, chandran2010}.

On the other hand, it has also been argued that the energy of Alfv\'{e}nic
turbulence does not cascade effectively to high-frequency parallel cyclotron
waves, but rather is transferred into low-frequency and highly oblique kinetic
Alfv\'{e}n waves. Then dissipation would still take place at the proton
gyroscale, but via Landau damping acting mostly on electrons
\citep{howes2008}.

In this letter, while benefiting from the high-resolution magnetic field data
and the detailed proton and alpha-particle velocity distribution functions
(VDFs) obtained \citep{marsch1982} simultaneously by the Helios~2 wave and
plasma experiments, we analysed typical solar wind data from a heliocentric
distance of about 0.7~AU. While studying them in detail we found that the
prominent kinetic features of the alpha particles and protons are closely
correlated with the power of the transverse high-frequency waves. The results
of our study place new empirical constraints on solar wind models and further
limit the theoretical assumptions involved. We will first present a
statistical analysis of the relevant plasma and field parameters, and then
discuss them in the light of kinetic plasma theory.

\section{Observations}

Here we analyse plasma and magnetic field data provided by the Helios~2
spacecraft in 1976. For the purpose of a solid statistical study, we selected
a week of continuous measurements, i.e., the days with DOY (day of year)
numbers 67 to 73, on which the spacecraft was at a solar distance of
about 0.7~AU.

\begin{figure}[t]
\centering \noindent\includegraphics[width=16pc]{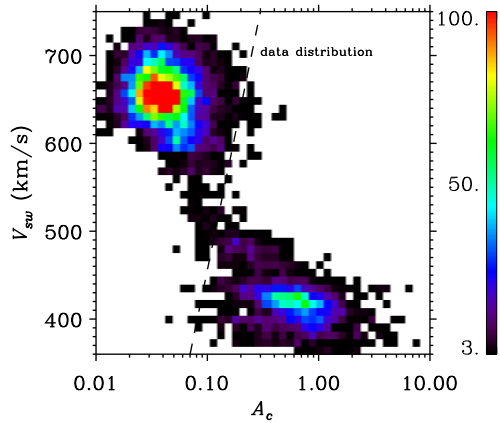}
\caption{Distributions of the number of data (presented in colour with the coding
indicated by the right-hand bar), which are  plotted as a function of the collision age,
$A_c$, and the solar wind speed, $V_{sw}$. The inclined dashed line separates
the data from fast and slow solar wind.} \label{fig.1}
\end{figure}

As is obvious from Figure~\ref{fig.1}, the selected set covers data from
measurements made in slow and fast solar wind. The slightly inclined dashed
line in that figure separates the data obtained in fast solar wind from those
in slow wind. As we would expect, the domain with a relatively high
number of collisions mainly corresponds to slow solar wind (with $A_c >
0.2$), and that with a low number of collisions corresponds to fast solar
wind which is weakly collisional (with $A_c < 0.1$). Here collisionality is
quantified by the so-called collision age \cite[see e.g.,][]{salem2003}, $A_c
=l/(V_\mathrm{sw}\tau_{\alpha p})$, where, $l$ is the distance from the sun
and $\tau_{\alpha p}$ is the collision-exchange time between helium ions and
protons.Moreover, for this data set the proton parallel
plasma beta, $\beta_{\parallel p}$, is comparatively small and mainly varies
between 0.1 and 0.7.

Because the solar wind speed is clearly higher than the Alfv\'en speed, we can
safely assume the so-called Taylor hypothesis to be valid. Therefore, the
spacecraft frequency of the magnetic field fluctuations (obtained from fast
Fourier transform) is simply given by, $2\pi f_\mathrm{sc}\simeq k
V_\mathrm{sw}$, where $V_\mathrm{sw}$ is the solar wind speed, and $k$ is the
wave number of the magnetic field fluctuations. Here, we deal with the
average wave power density, $\delta B^2$, which is obtained in the spacecraft
frame by integration over a frequency range that corresponds to the
wave-number range (0.01--0.1)$k_p$ (where $k_\rho = 1 / l_p$, $l_p$ is the
proton inertial length). Therefore, we are dealing with turbulent
fluctuations that still belong to the inertial range but are near the
dissipation range of the turbulence. We have chosen this frequency domain 
to avoid the inclusion of higher-frequency fluctuations that might stem
from local wave excitation (e.g., owing to a large temperature anisotropy or
high normalized ion differential speed) below but near the proton gyro-kinetic scale.
Consistently with our choice of this narrow frequency range, we define the
mean field (required as the reference value for the superposed fluctuations)
by the average of the full magnetic-field vector over the short time period
of about one hundred times $l_p/ V_\mathrm{sw}$.

\begin{figure}[t]
\centering \noindent\includegraphics[width=16pc]{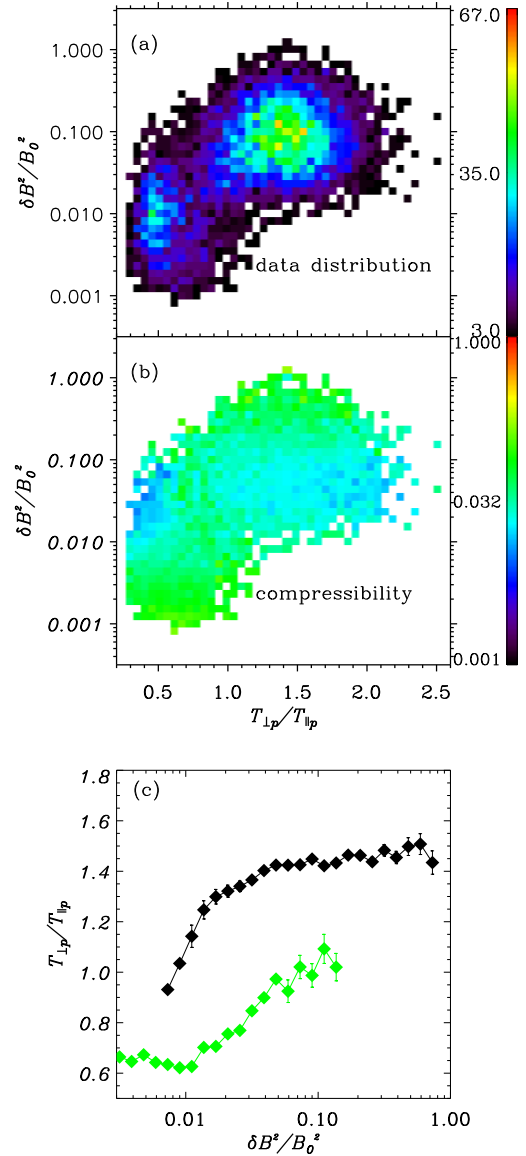}
\caption{Distributions (presented in colour with coding indicated by the
right-hand bars) of the number of our data (a) and the compressibility, $\delta
B_{\parallel}^2/(\delta B_{\perp}^2 + \delta B_{\parallel}^2)$ (b), plotted in 
the parameter plane of the normalized wave power versus the
proton temperature anisotropy, $T_{\perp p}/T_{\parallel p}$. (c)
Weighted mean value of the proton temperature anisotropy for $A_c <0.1$
(black), and $A_c>0.2$ (green) displayed versus normalized wave power. The
short vertical bars indicate the small uncertainties of the weighted mean values.}
\label{fig.2}
\end{figure}

In Figure~\ref{fig.2}~a the distribution of the normalized wave power,
$\delta B^{2}/B^{2}_{0}$, is plotted versus the proton temperature
anisotropy, $T_{\perp p}/T_{\parallel p}$. It turns out that the latter
parameter ranges between 0.4 and 2 and correlates positively with the
normalized wave power. According to Figure~\ref{fig.2}~b{\normalsize }, the
relative compressive component (coded in colour) of the fluctuations is
fairly small and not higher than $5\%$ of the overall wave amplitude, which
means that the fluctuations are mainly transverse and essentially
incompressible. It also appears that there is no striking correlation
between magnetic compressibility and normalized wave power. Unfortunately, 
we do not have the simultaneous small-scale velocity-fluctuations owing to
the lack of an adequate time resolution of the plasma experiment, and therefore we cannot
scrutinize the nature of the magnetic fluctuations or confirm by
polarization study whether they are Alfv\'enic or not. But presumably they
are Alfv\'{e}n waves, because these waves with periods longer than 40~s
are observed widely in fast solar wind \citep{tu1995} streams at all distances
between 0.3 and 1~AU.

In Figure~\ref{fig.2}c we plot the weighted mean value (expectation value) of the
proton temperature anisotropy as a function of the normalized wave power for
different solar wind regimes. In fast solar wind (i.e., where $A_c<0.1$,
black curve), the proton temperature anisotropy is below unity whenever the
value of the normalized wave power is below 0.01. However, above 0.01 the
temperature anisotropy steeply increases until the normalized wave power
reaches a value of about 0.1, where the temperature anisotropy tends to
saturate at a value of about 1.5. In contrast, when collisions are relatively
strong (i.e., where $A_c>0.2$, green curve), the proton temperature anisotropy
also increases with the wave amplitude but reaches lower values than those
when collisions are weak. The temperature anisotropy ranges between 0.6 and about 1.1,
while the normalized wave power varies between 0.01 and 0.1.

Our study shows that the proton temperature anisotropy correlates
positively with the normalized wave power.Moreover, a positive
correlation between the wave power and the proton temperature anisotropy was
found earlier by \citet{bourouaine2010}, but for another set of Helios data
restricted to fast solar wind and at various solar distances. However, the
Helios measurements selected for the present study were made at a fixed
heliocentric distance which is about 0.7 AU. Therefore, the new correlation
established here between the proton temperature anisotropy and normalized
wave power is certainly of local nature and not affected by radial
evolution.

According to Figure~\ref{fig.2}c, it seems that collisions play a major role
in reducing the effects waves have on the local heating of the protons.
Therefore, protons can become heated (if wave energy is available) more easily
in regions with relatively weak collisions than in regions where collisions
are relatively strong.

\begin{figure}[t]
\centering
\includegraphics[width=15.3pc]{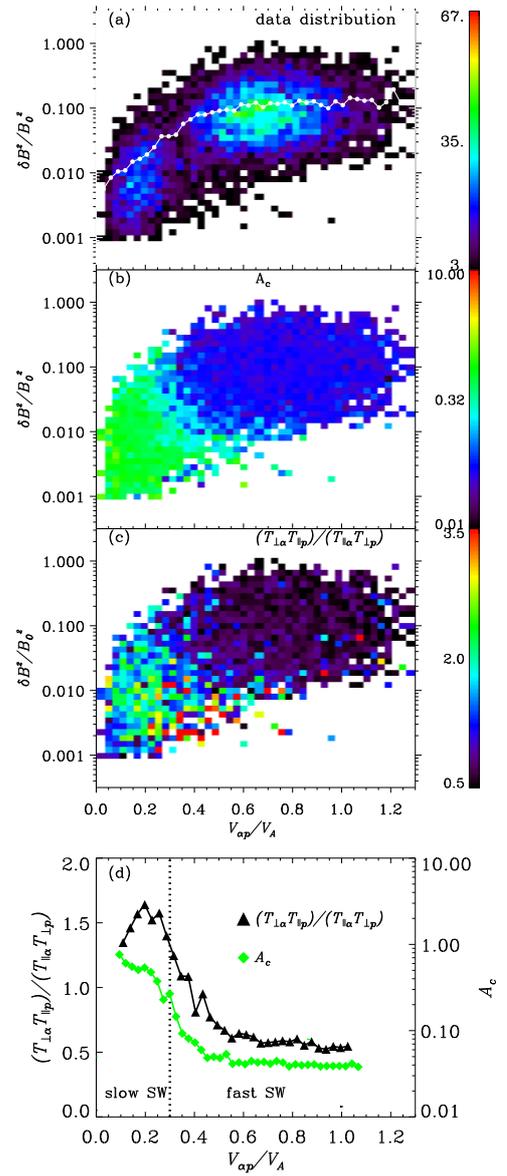}
\caption{Distributions (presented in colour with coding indicated by the
right-hand bars) of the number of our data (a), the collision age, $A_c$ (b) and
the ratio between the temperature anisotropies of alpha particles and
protons, $(T_{\perp \alpha} T_{\parallel p})/(T_{\parallel \alpha} T_{\perp
p})$, (c). All parameter are plotted in the plane of the normalized
wave-power, $\delta B^2/B_0^2$, versus normalized ion differential speed,
$V_{\alpha p}/V_\mathrm{A}$; (d) Weighted mean values of the
collisional age, $A_c$, and the ratio between the temperature anisotropies of
alpha particles and protons, $(T_{\perp \alpha} T_{\parallel
p})/(T_{\parallel \alpha} T_{\perp p})$, displayed versus normalized ion
differential speed, $V_{\alpha p}/V_\mathrm{A}$. The white dots in panel (a)
represent the weighted mean values. The vertical dotted line in panel (d)
separates the regimes of fast and slow solar wind.} 
\label{fig.3}
\end{figure}

Figure~\ref{fig.3}~a shows the data distribution of the normalized wave power
versus the normalized ion differential speed. It hardly exceeds unity, a
result which is consistent with the prediction of kinetic theory for a linear
plasma instability. The theory says that, whenever the ion differential speed
exceeds the Alfv\'en speed, the plasma should become unstable and excite
magnetosonic waves (see, e.g., \cite{li2000}). As \citet{bourouaine2011} found previously, 
the present study also indicates a positive
correlation between the normalized ion differential speed and the normalized
wave power (see the white curve, which represents the weighted mean values of
the normalized wave power shown in Figure~\ref{fig.3}~a. {\huge })

It is commonly believed (see, e.g., the review of \cite{marsch2006}) that low
collisional friction would permit a relatively high differential speed to
occur between the two main ion species in the solar wind. This notion is
confirmed by the results of our Figure~\ref{fig.3}~b, which shows that
higher values of the normalized ion differential speed correspond to lower
collision ages. In slow solar wind, when the value of the collision age is
higher than 0.2, the corresponding normalized differential speed is low
(i.e., $V_{\alpha p} /V_A < 0.3$), but it is higher than 0.3 for
comparatively low values of the collision age (as is indicated later in
Figure~\ref{fig.3}~d).

In Figure~\ref{fig.3}c the coloured pixels represent the ratio of the
alpha-particle-to-proton temperature anisotropy, $(T_{\perp\alpha}
T_{\parallel p})/(T_{\perp p} T_{\parallel \alpha})$, plotted as a function
of the relative ion differential speed and the normalized wave power.
Interestingly, this figure clearly shows that when $V_{\alpha p} /V_A \le
0.4$, the temperature anisotropy of the alpha particles, $T_{\perp
\alpha}/T_{\parallel \alpha}$, is higher than the anisotropy of the protons,
$T_{\perp p}/T_{\parallel p}$. However, the ratio of the ion temperature
anisotropies tends to decrease to lower values of about 0.6 when $V_{\alpha
p} /V_\mathrm{A} > 0.6$, as is quantitatively shown in Figure ~\ref{fig.3}~d.

The curve in Figure~\ref{fig.3}d, which represents the weighted mean value of
the ratio of the ion temperature anisotropies (black symbols), clearly
indicates that alpha particles are preferentially heated (perpendicularly to
the mean magnetic field) with respect to the protons whenever $V_{\alpha p}
/V_\mathrm{A} \le 0.4$, and this is true even for a relatively low wave energy
(indicated by the white curve in Figure~\ref{fig.3}a) and at high collision
rates (green symbols).

One would expect that the plasma tends to thermal equilibrium, in
coincidence with the lowest values of the normalized ion differential flow speed, if a relatively 
high collision rate. Then the temperature ratios of the
ion species should also be near unity. However, observationally it seems that
preferential perpendicular heating of the alpha particles with respect to the
protons can persist even in regions where collision rates are high (with
$V_{\alpha p} /V_A \le 0.4$). This is possible because a wave-related local
ion heating mechanism may be acting on time scales much lower than the long
cumulative collision time. Such a fast wave-heating mechanism can drive the
plasma far away from thermal equilibrium, and therefore may cause a significant
ion temperature anisotropy, because on the other side collisions are not fast
enough to enforce thermal equilibrium.

We found in our previous paper (\citet{bourouaine2011}) that the helium ion
abundance for this selected data set varies mainly between 0.02 and 0.04. The
helium abundance does not show a clear dependence on the normalized
differential ion speed. However, we showed that there is an anti-correlation between the alpha-to-proton
temperature ratio and the helium abundance at a fixed $V_{\alpha p} /V_\mathrm{A}$.

\begin{figure}
\centering
\includegraphics[width=21pc]{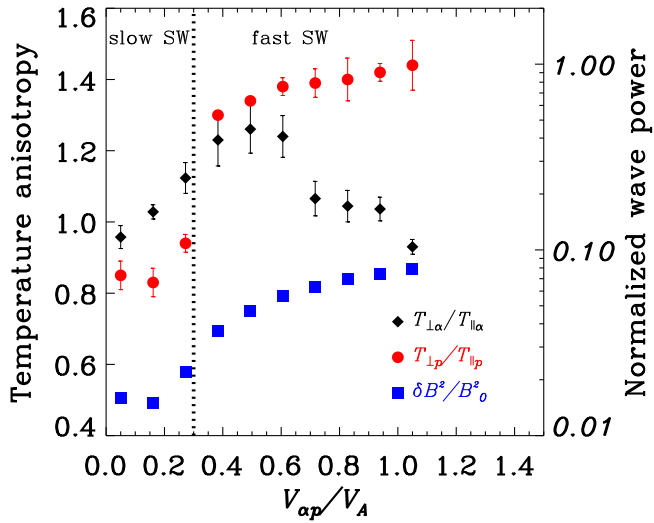}
\caption{Left ordinate: The mean values of the temperature anisotropy of the
protons, $T_{\perp p}/T_{\parallel p }$ (red dots), and of the alpha
particles, $T_{\perp \alpha}/T_{\parallel \alpha}$ (black diamonds). Right
ordinate: Mean normalized wave power (blue squares). The bars indicate the
uncertainties of the mean values. All quantities are plotted in bins versus
the relative ion differential speed, $V_{\alpha p}/V_\mathrm{A}$. The
vertical dotted line separates fast from slow solar wind regimes.}
\label{fig.4}
\end{figure}

Figure~\ref{fig.4} is a plot of the mean values of the alpha-particle
temperature anisotropy, $T_{\perp \alpha}/T_{\parallel \alpha}$, the proton
temperature anisotropy, $T_{\perp p}/T_{\parallel p}$, and the mean relative
wave power versus $V_{\alpha p}/V_\mathrm{A}$. The mean values are obtained
by averaging the data within bins of a width $\Delta (V_{\alpha
p}/V_\mathrm{A}) = 0.1$, and the vertical error bars indicate the related
uncertainties of the mean values.

As mentioned above, it turns out that the proton temperature ratio is
strictly correlated with the normalized wave power, which indicates perpendicular
heating of the protons whenever the power of the transverse fluctuations is
enhanced.

An interesting behaviour of the temperature ratio of the alpha particles can
be inferred from Figure~\ref{fig.4}. This ratio increases with increasing
normalized wave power as long as the normalized ion differential speed stays
below about 0.5. Beyond this value of $V_{\alpha p}/V_\mathrm{A}$, the
alpha-particle temperature ratio becomes roughly constant, until $V_{\alpha
p}/V_\mathrm{A}$ exceeds a value of about 0.7, but then it decreases towards
a value below unity when $V_{\alpha p}/V_\mathrm{A}$ reaches one. In the slow
solar wind region, where the collisions are expected to be relatively high,
the proton temperature anisotropy is ranging between 0.8 and 1, however, the
alpha temperature anisotropy is not in the same range but higher, and varies
between 0.9 and 1.2.

Most likely strong collisionality at low $V_{\alpha p}/V_\mathrm{A}$ tends
to isotropize the alphas, whereas the weakening of the resonance at higher
alpha/proton speeds leads to a decreasing of the alpha anisotropy, as in
\citet{gary2005}.

The monotonic increase of the proton temperature anisotropy with the
normalized differential ion speed appears to be consistent with the findings
of \citet{kasper2008} and with the ACE observations of \citet{gary2005}).
However, there is clear difference between the trend of the alpha temperature
anisotropy in Figure~\ref{fig.4} and the results found by \citet{kasper2008}.
In their paper, the alpha temperature anisotropy reaches a minimum value at
$V_{\alpha p} \sim 0.5 V_\mathrm{A}$, whereas our Figure~\ref{fig.3} shows
that the alpha temperature anisotropy reaches its maximum value when
$V_{\alpha p} \sim 0.5 V_\mathrm{A}$. Moreover, the results of
\citet{gary2005} show that the average alpha temperature anisotropy is
monotonically decreasing with increasing alpha/proton relative speed.

Unlike what \citet{kasper2008} and
\citet{gary2005} found previously, our Figure~\ref{fig.4} shows that the perpendicular heating
is reduced when $V_{\alpha p}/V_A$ is near zero. This reduced ion heating
corresponds to an observed concurrent decrease in the wave power of the
transverse waves. This results would be expected when the heating
ultimately rests with the energy in those waves. If $V_{\alpha p}/V_A$ is
near zero, we expect that the alpha-particle temperature ratio
increases resulting in a strong perpendicular ion heating for sufficient wave
power. However, although the wave power is empirically found to be weak when
the normalized ion differential speed is low, the alpha particles are still heated
perpendicularly much more than the protons. For both ion species the
interaction with transverse waves is expected to work against the radial
trend caused by the solar wind expansion in a magnetic mirror, which tends to
build up a much larger parallel than perpendicular temperature, and accordingly a
strong fire-hose-type anisotropy.

\section{Discussion and conclusions}

If we assume that the transverse fluctuations provide the energy input for the
observed preferential perpendicular heating of the alpha particles, then
their temperature-ratio profile as given in Figure~\ref{fig.4} could be a
signature of cyclotron-wave heating of the alpha particles.

It has been claimed that the alpha particles can only be heated through
ion-cyclotron wave dissipation if the differential speed between
alpha-particles and protons is approximately less than $0.5V_A$ \citep[see
e.g.,][]{gary2005}. Therefore, alpha-particles can be heated in the
perpendicular direction as long as they stay in resonance with the
ion-cyclotron waves. Moreover, the wave energy is an important input
parameter that controls the heating of alpha-particles. We expect that high
wave energy can cause strong perpendicular heating of those alpha
particles that are in resonance with ion-cyclotron waves.

Figure~\ref{fig.4} shows that the perpendicular heating is reduced when
$V_{\alpha p}/V_\mathrm{A}$ is near zero. This reduced ion heating
corresponds to a decrease in the observed wave power of the transverse
fluctuations, a result that is expected because the potential for heating
ultimately rests with the energy contained in those waves. If $V_{\alpha
p}/V_\mathrm{A}$ is near zero, we expect that the alpha-particle
temperature ratio increases, resulting in strong perpendicular ion
heating for sufficient wave power. However, although the wave power is found
empirically to be weak when the normalized ion differential speed is small, the alpha
particles can still be perpendicularly heated much more than the protons.
Yet, for both ion species the interaction with transverse fluctuations works
against the trend caused by the solar wind expansion in a magnetic-field
mirror configuration, which tends to build up a high parallel temperature
anisotropy. Furthermore, Coulomb collisions tend to thermalize the solar wind
plasma and to reduce the differential speed between alpha particles and
protons. But collisions are effective in removing the non-thermal ion
features merely in the comparatively cold and dense slow solar wind, in which
the collision age is found to be high and the average wave power observed to
be weak.

Another possible scenario that may occur as well is that the
long-wavelength and high-amplitude fluctuations of the inertial-range
turbulence may stochastically heat \citep{chandran2010} the ions in the
perpendicular direction with respect to the background magnetic field (or
non-resonantly drive a slowly varying $T_\perp/T_{\parallel}$ on both ion
species). The presence of this anisotropy can give rise to
Alfv\'en-cyclotron instabilities, which lead to the growth of relatively
high-frequency modes (with frequencies $\omega \sim \Omega_p$, where
$\Omega_p$ is the proton cyclotron frequency). In other words, the ion
temperature anisotropies could first be caused by low-frequency fluctuations
in the inertial range, and then this thermal energy may be exchanged between
ions and waves at the proton kinetic scale.

%

\end{document}